# Dir-MUSIC Algorithm for DOA Estimation of Partial Discharge Based on Signal Strength represented by Antenna Gain Array Manifold

Wencong Xu[1], Yandong Li[2], Bingshu Chen[2], Yue Hu[2*], Jianxu Li[1], Zijing Zeng[2]

[1] Department of automation, Shanghai Jiao Tong University, No.800 Road Dongchuan, Shanghai, China
[2] Department of Electrical Engineering, Shanghai Jiao Tong University, No.800 Road Dongchuan, Shanghai, China

**Abstract:** Inspection robots are widely used in the field of smart grid monitoring in substations, and partial discharge (PD) is an important sign of the insulation state of equipments. PD direction of arrival (DOA) algorithms using conventional beamforming and time difference of arrival (TDOA) require large-scale antenna arrays and high computational complexity, which make them difficult to implement on inspection robots. To address this problem, a novel directional multiple signal classification (Dir-MUSIC) algorithm for PD direction finding based on signal strength is proposed, and a miniaturized directional spiral antenna circular array is designed in this paper. First, the Dir-MUSIC algorithm is derived based on the array manifold characteristics. This method uses strength intensity information rather than the TDOA information, which could reduce the computational difficulty and the requirement of array size. Second, the effects of signal-to-noise ratio (SNR) and array manifold error on the performance of the algorithm are discussed through simulations in detail. Then according to the positioning requirements, the antenna array and its arrangement are developed, optimized, and simulation results suggested that the algorithm has reliable direction-finding performance in the form of 6 elements. Finally, the effectiveness of the algorithm is tested by using the designed spiral circular array in real scenarios. The experimental results show that the PD direction-finding error is 3.39°, which can meet the need for Partial discharge DOA estimation using inspection robots in substations.

## 1. Introduction

Partial discharge [1] is one of the main causes of power equipment insulation failure. Ultra high frequency (UHF) detection method [2] is widely used due to its strong anti-interference ability and high sensitivity. To improve detection efficiency and reduce detection cost, some scholars have proposed to use the omnidirectional UHF sensor array to detect and locate the PD signal in the substation area, which gets fruitful results [3-10].

The status inspection of power equipment is one of the key contents of substation operation and maintenance. The development of smart substations puts forward higher requirements for substation inspection. An intelligent inspection robot [11-13] is an effective way to replace manual inspection and has broad application prospects. Existing inspection robots generally use infrared and high-definition cameras to perform temperature and status detection. However, limited by the implementation conditions of the small and light-weight sensor array, currently, no inspection robot can effectively detect and locate PD. Therefore, it is of high engineering application value to research a lightweight, small-sized, high-gain antenna array that can be mounted on inspection robots.

To estimate the PD direction, spatial spectrum estimation in the field of array signal processing obtains the spatial and temporal observation data of the source space by sampling the spatially distributed field signals. It has the advantage of super-resolution and beyond the Rayleigh limit, providing a theory for the study of partial discharge positioning algorithms. In the 1970s, Doctor Schmidt from Stanford University proposed the well-known Multiple Signal Classification (MUSIC) algorithm [14-16], which detects direction based on the orthogonality of signal feature subspace and noise feature subspace. Afterward, based on the MUSIC algorithm, a large number of improved algorithms were derived [17-22]. At present, there are few related types of research on spatial spectrum estimation in the field of partial discharge detection, and they mainly focus on the optimization of ultrasonic array arrangement and algorithm improvement [23-24]. Also, the existing algorithms have the following limitations: (1) The essence of the algorithm is to do correlation calculation using the phase difference caused by the wave path difference of the array signal. If the collected signal is a UHF signal, the requirement for the sampling rate of the device is very high. (2) The calculation of the phase difference based on the delay difference depends on the signal frequency, so the algorithm is only suitable for narrowband signals. Because the PD signal is a wideband signal, the focusing algorithm needs to be used to convert the received data. However, the introduction of the focusing method does not only increases the amount of calculation, and some errors are also introduced by the conversion. (3) The element spacing of the array cannot be greater than half of the wavelength of the incident wave, otherwise, the angle will be blurred.

Based on the superiority of the spatial spectrum algorithm and the possibility of the algorithm's migration from phase information to signal strength information, this paper focused on the directional MUSIC (Dir-MUSIC) algorithm research based on signal strength information and carried out the research work of signal model derivation, Dir-MUSIC algorithm performance research, PD sensor array development, and algorithm effectiveness verification. Compared to the phase information acquirement, the signal strength information acquirement can greatly reduce the requirement of the sampling rate and computational complexity. We show through simulation and experiment results that the proposed Dir-MUSIC algorithm works well under different conditions of SNR and slight array manifold error. The average PD localization error is 3.39° through experimental results with only 6 elements of the spiral antenna circular array, which proves the engineering practical value of the proposed method for substation inspection robots.

The rest of the paper proceeds as follows: Section 2 introduces an analytical expression of the antenna pattern and derives the Dir-MUSIC algorithm using antenna gain array manifold and signal strength information. In section 3, simulation results under the condition of SNR and array manifold error are provided. Section 4 presents the PD localization experimental results with the designed directional spiral antenna circular array, and the conclusions are outlined in section 5.

## 2. Dir-MUSIC Algorithm

The traditional MUSIC algorithm is only suitable for narrowband signals. The reason is that the array manifold depends on the signal frequency. If directional antennas are used instead of omnidirectional antennas and the strength array manifold is formed based on the array pattern, the limitation of narrowband will be eliminated, which will significantly expand the range of application of the MUSIC algorithm. Suppose the antenna array is a uniform circular array of N elements (directional antenna), and the noise is Gaussian white noise with a mean value of zero, which is independent of the signal. Also, suppose the array is in the far-field area of the source radiation, that is, what the antenna array receives from the source is a plane wave.

Any continuous function can be approximated by a linear combination of Gaussian functions. Based on the actual antenna pattern, a linear combination of 3 Gaussian functions is enough to describe the two-dimensional antenna pattern function, as shown in equation (1):

$$g(\theta) = a_1 e^{-\frac{(\theta-b_1)^2}{c_1^2}} + a_2 e^{-\frac{(\theta-b_2)^2}{c_2^2}} + a_3 e^{-\frac{(\theta-b_3)^2}{c_3^2}} \quad (1)$$

where a1=0.5255，b1=218.1，c1 =51.73，a2=0.3405，b2=304.8，c2 =41，a3 = 0.6251，b3 =156.1，c3 =109.1. There parameters are calculated according to the mesured antenna pattern. The normalized atennas gain array manifold(AGAR) with 6 elements is shown in Fig. 1.

Suppose the source signal is (t), which is a broadband signal, and the direction of the incoming wave is θ, then the gain matrix of each element for the incoming wave in the direction is shown in (2):

$$G = [g_1(\theta), g_2(\theta), \cdots, g_N(\theta)]^T, \quad (2)$$

where we propose an assumption that the response of each element is approximately equal in the same direction and at a different frequency in the frequency band this paper focus on.

The response of each element to the source is shown in (3):

$$X = GS = [g_1(\theta)s(t), g_2(\theta)s(t), \cdots, g_N(\theta)s(t)]^T. \quad (3)$$

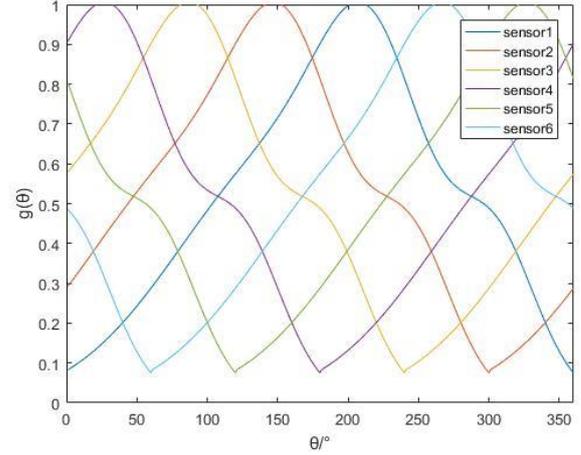

*Fig. 1. Normalized AGAR*

Considering the noise effect, the matrix form is shown in (4):

$$X = GS + N, \quad (4)$$

where $S = s(t)$，$N = [n_1(t), n_2(t), \cdots, n_N(t)]^T$, and $n_i(t)$ is independent identically Gaussian distributed

Perform covariance processing on the array output X to obtain the covariance matrix $R_X$:

$$\begin{aligned}R_x &= E(XX^T) \\ &= E[(GS + N)(GS + N)^T] \\ &= GE(SS^T)G^T + 0 + 0 + E(NN^T) \\ &= GR_s G^T + R_N,\end{aligned} \quad (5)$$

In (5), $R_S = E(SS^T)$ is the correlation matrix of the signal, $R_N = E(NN^T) = \sigma^2 I$ is the correlation matrix of the noise.

Suppose the direction angle of the incoming wave relative to the element 1 is $\theta_1$, then the array gain matrix G of a uniform circular array of N elements can be obtained, that is, the strength array manifold is shown in equation (6):

$$G = \begin{bmatrix} g_1(\theta) \\ g_2(\theta) \\ \vdots \\ g_N(\theta) \end{bmatrix} = \begin{bmatrix} a_1 e^{-\frac{(\theta_1-b_1)^2}{c_1^2}} + a_2 e^{-\frac{(\theta_1-b_2)^2}{c_2^2}} + a_3 e^{-\frac{(\theta_1-b_3)^2}{c_3^2}} \\ a_1 e^{-\frac{(\theta_1+\frac{360}{N}-b_1)^2}{c_1^2}} + a_2 e^{-\frac{(\theta_1+\frac{360}{N}-b_2)^2}{c_2^2}} + a_3 e^{-\frac{(\theta_1+\frac{360}{N}-b_3)^2}{c_3^2}} \\ \vdots \\ a_1 e^{-\frac{(\theta_1+\frac{360}{N}(N-1)-b_1)^2}{c_1^2}} + a_2 e^{-\frac{(\theta_1+\frac{360}{N}(N-1)-b_2)^2}{c_2^2}} + a_3 e^{-\frac{(\theta_1+\frac{360}{N}(N-1)-b_3)^2}{c_3^2}} \end{bmatrix} \quad (6)$$

that quation (6) suggests hat for a single source, if the direction of the incoming wave is determined, it is impossible for each element of the G matrix to be 0, i.e. the strength array manifold matrix is full rank. Based on the previous assumption, $R_S$ is the square of the modulus of the acquired signal, i.e. the signal power, whose value is greater than 0; $R_N$ is the correlation matrix of noise, then it can be derived that $R_X$ is a full-rank matrix, and $rank(R_x) = N$. Also

$$R_X = E(XX^T) = E[(XX^T)^T] = R_X^T, \quad (7)$$

therefore, $R_X$ is a Hethe rmite matrix. According to the characteristics of Hermite matrix, its eigenvalaues are all real numbers. Then, because $R_S$ is positive definite, $R_X$ is a positive definite matrix. Suppose the eigenvalus of $R_X$ are $\lambda_1, \lambda_2, \cdots, \lambda_N$, and the corresponding feature vectors are $\upsilon_1, \cdots, \upsilon_N$. Since $R_X$ is a Hermite matrix, whose eigenvectors are orthogonal to each other, that is:

$$v_i^T v_j = 0, i \neq j \quad (8)$$

Because there is only 1 signal source, there is 1 eigenvalue related to the signal, which is respectively equal to the sum of the eigenvalue of $GR_sG^T$ and $\sigma^2$, and the remaining N-1 eigenvalues are related to the noise, and their values should be $\sigma^2$ in theory. Thus, the N-1 eigenvalues related to the noise are relatively small. Using this characteristic, sort the eigenvalues of $R_X$ from largest to smallest.

$$\lambda_1 \geq \lambda_2 \geq \cdots \geq \lambda_N \quad (9)$$

Let $\nu_i$ be the eigenvector corresponding to $\lambda_i$, it can be derived according to the definition that:

$$R_X\nu_i = \lambda_i\nu_i \quad (10)$$

Suppose that $\lambda_i = \sigma^2$ is the smaller N-1 eigenvalues, and the corresponding eigenvectors are $\lambda_j$ respectively, and it can be obtained that:

$$R_X\nu_j = \lambda_j\nu_j \quad (11)$$

Then it can be obtained that:

$$(GR_sG^T + \sigma^2 I)\nu_j - \sigma^2\nu_j = 0 \quad (12)$$

It can be obtained by some simplification that:

$$G^T\nu_j = 0, j = 2, \ldots, N \quad (13)$$

Equation (13) suggests that the column vector of matrix G and the eigenvector corresponding to the noise space are orthogonal and because matrix G contains the information of incoming wave, equation (13) can be used as the basis for finding the direction of incoming wave. Use N-1 eigenvectors to construct a noise matrix:

$$E_n = [\nu_{1+1}, \nu_{2+1}, \cdots, \nu_N] \quad (14)$$

Let the space spectrum be (15):

$$P_{mu}(\theta) = \frac{1}{g^T(\theta)E_nE_n^Tg(\theta)} = \frac{1}{\|E_n^Tg(\theta)\|^2} \quad (15)$$

By traversing the angle $\theta$, the spatial spectrum changes. When the maximum value of $P_{mu}(\theta)$ is taken, it indicates that $g(\theta)$ is orthogonal to the noise matrix. At this time, $\theta$ is the estimated direction angle of the incoming wave.

## 3. Simulation Research on Algorithm Performance

### 3.1 Relation Between Incoming Wave Direction Esimation and SNR

To explore the performance of the algorithm under different SNR, the Dir-MUSIC algorithm was simulated and tested based on the MATLAB platform to verify its reliability. The simulation process is shown in Fig. 2.

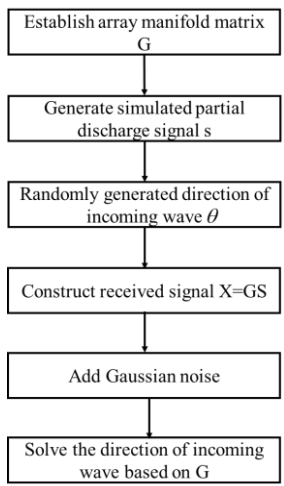

*Fig. 2 Simulation flow chart of Dir-MUSIC*

Suppose the number of element N is 6, SNR is 10 dB. First, the strength array manifold was constructed based on the array element pattern, as shown in Fig. 1. Then, 3600 incoming wave directions were randomly generated in the range of the circle [1°, 360°], and 3600 simulations were performed. The accuracy of the incoming wave direction estimation is defined as: if the error between the estimated angle and the actual angle is less than 2°, the positioning is successful; the accuracy is equal to the ratio of the number of successful positioning to the total number of simulations. The double exponential oscillation attenuation function was used to simulate the PD pulse signal, the received signal of the array after adding noise is shown in Fig. 3.

The accuracy of the incoming wave direction estimation for different SNR is shown as Table 1.

**Table 1** Direction finding accuracy under different SNR

| SNR | 10 | 5 | 0 | -5 | -10 |
|---|---|---|---|---|---|
| Accuracy | 100% | 100% | 100% | 99.17% | 72.78% |

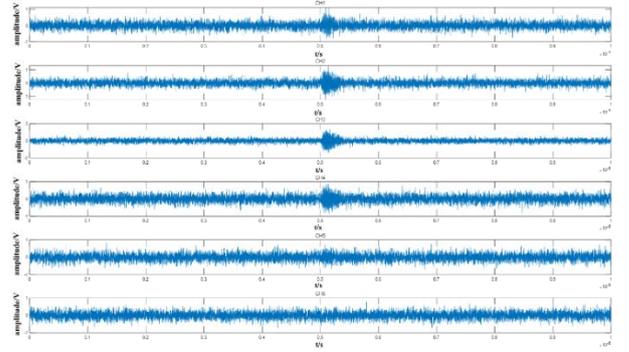

*Fig. 3 Waveform of 6-channel PD pulse after noise addition (SNR = -10)*

It can be seen from Table 1 that the proposed algorithm had a strong anti-interference effect on Gaussian white noise. When the SNR dropped from 10 to -5, the estimation accuracy of the direction of arrival could still reach 99.17%. When the SNR was -10, as shown in Fig. 3, the PD pulses of channel 4 and channel 5 had been completely covered by noise, and the PD pulses of channel 1 and channel 4 were not obvious. At this time, the accuracy rate was still up to 72.78%. The figure below shows the angle estimation error for different incoming wave directions when the SNR was -10.

As shown in Fig.4, the upper limit of the error distribution was 8, the lower limit was -7, the mean was 0.0583, and the variance was 2.2291. The distribution of angle estimation errors was relatively concentrated, which met the positioning requirements.

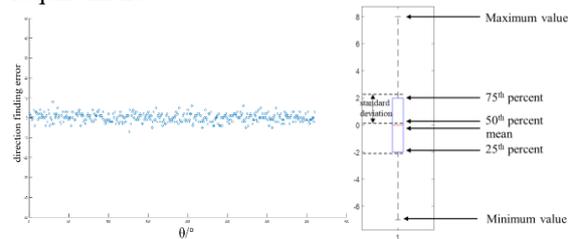

*(a) Error scatter plot     (b) Error box diagram*
*Fig. 4 Direction finding error in different directions(SNR = - 10)*

## 3.2. Research on Direction-Finding Performance under the Condition of Array Manifold Error

When the array is an ideal model, the Dir-MUSIC algorithm has excellent direction-finding performance. However, when there exist errors in the array manifold, the performance of the algorithm will rapidly decrease or even fail. In practical applications, due to the antenna manufacturing process, signal acquisition channels, working environment and other factors, the amplitude error (mainly between the actual G and the estimated G) greatly affects the direction-finding performance of the proposed algorithm. When the array has an amplitude error, the array manifold for 6 antennas becomes:

$$G' = [g_1(\theta) + \Delta g_1, g_2(\theta) + \Delta g_2, \ldots, g_6(\theta) + \Delta g_6]^T \quad (16)$$

Then the actual received data is:
$$X = G'S + N \quad (17)$$

In the subsequent spectrum search process, the estimated G is still used as the array manifold from calculation, which caused errors in direction-finding. This section will explore the impact of different amplitude errors on direction-finding performance. In this section, random errors which are uniform distributed are added to the array manifold. Respectively, U(-0.1, 0.1), U(-0.075, 0.075), U(-0.05, 0.05) and U(-0.025, 0.025) were added to form the array manifold G' with errors, and the received signal X was generated using G', and the direction finding was performed based on the array manifold G without errors. In the case of SNR of 10, 3600 Monte Carlo simulations were performed, and the simulation results are shown in Fig. 5.

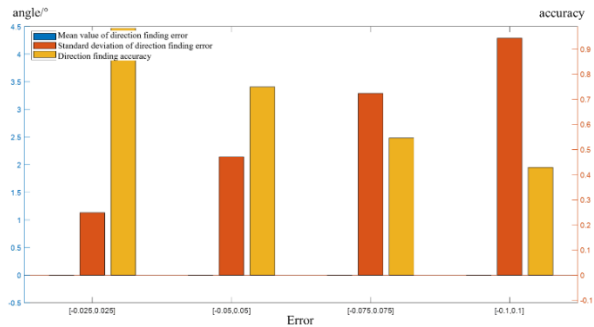

***Fig. 5*** *Direction finding results under different array manifold errors*

It can be seen from Fig. 5 that when the error increased from U(-0.025, 0.025) to U(-0.1, 0.1), the direction-finding accuracy decreased from 98.30% to 42.89%, and the standard deviation of the error increased from 1.13° to 4.29°. However, the standard deviation of the error was about 0, which did not change with the amplitude error. The mean value was almost unchanged because the added error was a random error that conformed to the uniform distribution law. This regular error caused the spatial spectrum peak to move around the true peak, and this movement was affected by the error. When the error was a uniformly distributed error, the size of the spatial spectrum peak movement conformed the uniform distribution law. The greater the error, the greater the position of the movement. Therefore, the mean value of the direction-finding error did not change with the size of the error added by the array manifold, and the standard deviation of the direction-finding error increased with the increase of the error. When the error of U(-0.1, 0.1) was added to the normalized array manifold, the standard deviation of the direction-finding error was only 4.29°, which shows that the proposed algorithm had better robustness to the amplitude error of the array.

## 3.3 Relation Between Incoming Wave Direction Estimation and Number of Array Element

The Direction-finding performance is also strongly related to the number of array elements. In traditional spatial spectrum estimation, the number of array elements determines the array aperture. The larger the number of array elements, the larger the array aperture, and its direction-finding performance gets better. On the premise that the pattern of a single element is determined, the number of elements determines the array manifold. Due to the miniaturization requirements of the array, the radius of the array cannot exceed the radius R of the chassis of the inspection robot, which is about 20 cm. Assuming that the radius R of the uniform circular array is constant, then the number of array elements affects the distance between adjacent array elements, and different array element distances make the steering vector different, as shown in Fig. 6.

In Fig. 6(a), the 4-element array (blue) has an element pitch angle of 90°, and the 8-element array (black) has an element pitch angle of 45°. The corresponding array manifolds are

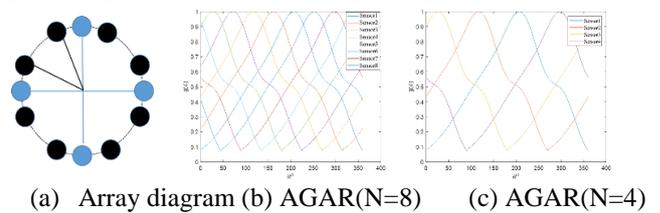

(a) Array diagram  (b) AGAR(N=8)  (c) AGAR(N=4)

***Fig. 6*** *Array diagram and AGAR with different number of elements*

shown in Fig. 6(b) and Fig. 6(c) respectively. The more the number of array elements, the greater the data redundancy of the signal receiving matrix and the stronger the gain performance of the array, theoretically, its direction-finding performance gets better. However, due to the large size of the UHF antenna, if the array elements are too close, the mutual coupling between the array elements will affect the antenna performance. Therefore, it is necessary to select the optimal number of array elements to ensure both the array size and the array element performance while ensuring the direction-finding performance.

Set the number of array elements to 1, 2, 4, 6, 8, and 10 respectively, the channel amplitude error to U(-0.05, 0.05), the signal-to-noise ratio to 10, and the direction of the incoming wave was randomly generated within a 360° circle. 3600 Monte Carlo simulations was carried out and the results are as shown in Fig .7.

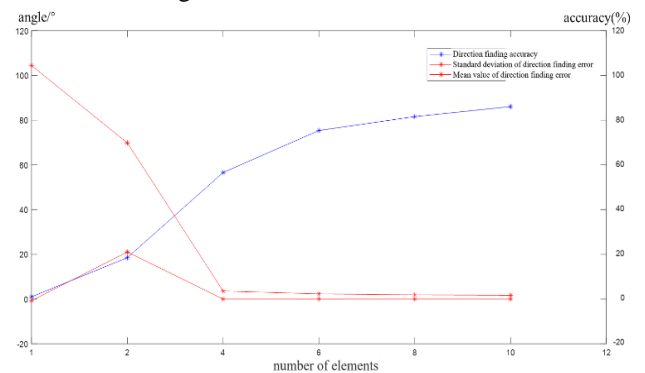

*Fig. 7 Direction finding performance under different array elements*

The results suggest that the accuracy of direction-finding is very sensitive to the number of elements. When the number of array elements was increased from 1 to 10, the direction-finding accuracy continuously increased from 1% to 86.06%. This was because when the number of array elements was too small, taking a single-element array as an example, the number of elements was equal to the number of incident wave, so the covariance matrix of the received data could not be decomposed into signal space and noise space, and the orthogonality of the two spaces could not be used for direction finding, resulting in the failure of MUSIC algorithm. With the increase in the number of array elements, the redundancy of the received data increased. The covariance matrix of the data could be decomposed into signal space formed by the eigenvectors corresponding to 1 signal and noise space formed by the eigenvectors corresponding to (number of array elements-1) signals. Multiple linearly independent noise vectors could make the noise space more robust to finite data length and amplitude errors. At the same time, when the number of elements increased, the gain performance of the array was also significantly improved, so the estimation of the signal space using received data was more accurately, and the direction-finding accuracy increased with the increase of the number of array elements.

As the number of elements increased, the speed of direction-finding accuracy slowed down. When the number of elements increased from 2 to 4, the accuracy of direction finding increased by 205.8%; when the number of elements increased from 4 to 6, the accuracy of direction finding increased by 33.2%. But when the number of elements increased from 6 to 8 and from 8 to 10, the accuracy of direction finding only increased by 11.35% and 5.4% respectively. This was because when the number of array elements was increased to 6, the eigenvectors corresponding to the noise space from the covariance matrix decomposition could already approximate the real noise space while maintaining a high gain of the array, so the increase in the number of array elements had a weaker and weaker effect on the increase of direction-finding accuracy.

When the number of array elements was too large, on the one hand, it would cause difficulties in data collection and processing. The data amount of the covariance matrix of the received data was proportional to the square of the number of elements. Too many elements would lead to an increase in the amount of calculation during the decomposition of the covariance matrix, which affected the real-time performance of the algorithm. On the other hand, due to the limitation of the maximum radius of the array, too many array elements would reduce the distance between the array elements, increase the mutual coupling error of the array, and affect the performance of the array. Therefore, generally considering the direction-finding accuracy, the statistical distribution law of the direction-finding error, the ability of data acquisition processing, and the array performance, the 6-element uniform circular array layout was the optimal formation.

It can be seen from the pattern of a single array element that when the angle deviated greatly from the front of the antenna, the gain was attenuated rapidly. Therefore, when waves arrived in these directions, the signals sensed by the antenna were almost completely submerged in noise, so the impact would be relatively small if this part of the data pair was discarded. Also, in practical applications, 4-channel data acquisition cards or digital oscilloscopes are often used. Therefore, the algorithm performance of the 4-element array was studied according to the element interval of the 6-element uniform circular array. The SNRs were set respectively at 10, 5, 0, -5, -10 to explore the direction-finding accuracy of the algorithm, and the results are shown in Table 2. It can be seen from Table 2 that when the SNR was high, both the 4-array element and the 6-array element could maintain high direction-finding accuracy. When the SNR was reduced to -10, the direction-finding accuracy of the 4-array element dropped rapidly to 47.67%. Therefore, in practical applications, in an environment with a high SNR, a 4-element array can be used for direction finding.

**Table 2** Direction finding accuracy under different SNR

| SNR | 10 | 5 | 0 | -5 | -10 |
|---|---|---|---|---|---|
| Accuracy | 100% | 100% | 98.17% | 85.62% | 47.67% |

## 4. Experimental System Construction and Positioning Test

### 4.1. Experimental System Construction

UHF sensors are used to convert electromagnetic signals excited by PD into electrical signals. It is the first and most important step in the detection and positioning of PDs. The performance of the sensor directly affects the detection effect and positioning accuracy. First, the electromagnetic pulse excited by PD is a broadband signal, so the sensor needs to have broadband characteristics. Secondly, based on the Dir-MUSIC algorithm proposed in this paper, the sensor needs to have good directional and gain characteristics. Finally, the used sensor array needs to meet the miniaturization requirements. Based on the above design requirements of the sensors, using the CST simulation software, the sensor array was designed as shown in Fig. 8.

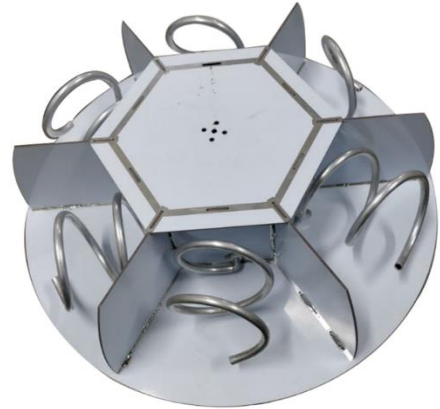

*Fig. 8 Physical picture of directional spiral antenna array*

The pattern of the directional helical antenna array is very important to the direction-finding performance of the algorithm. The microwave anechoic chamber can shield external electromagnetic interference. The absorbing material inside the anechoic chamber has a good effect on electromagnetic wave reception and can effectively suppress refraction and reflection. The detection system consisted of a shielded shell with a length of 19 meters, a width of 8.5 meters, and a height of 8.5 meters, and N5225A vector network analyzer, a reflecting surface, a turntable, a power amplifier, a low noise amplifier, and a feed source. The vector network analyzer was connected to the emitting feed source through the power amplifier. The emitted electromagnetic wave was converted into a plane wave by the reflecting surface and propagated to the antenna to be tested. The antenna to be tested was connected to the vector network analyzer through the low-noise amplifier for data analysis. The measurement of the XOY surface pattern of the antenna was completed by controlling the rotation of the turntable via an industrial PC. The specific detection system is shown in Fig. 9.

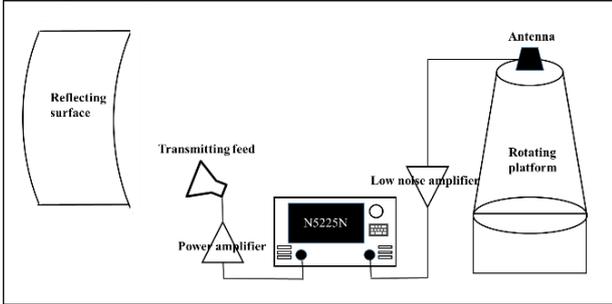

*Fig. 9 Diagram of microwave anechoic chamber detection system*

The array detection platform is shown in Fig. 10. The electromagnetic wave was emitted from the horn antenna, and the electromagnetic wave formed a plane wave after passing through the wave reflecting surface. The antenna received the plane wave and transmitted the received signal to the vector network analyzer through the coaxial cable connected to the antenna for calculating the gain of the antenna. The XOY surface pattern of the antenna was then measured by controlling the turntable. The measured pattern was the gain relative to the broadband double-ridged horn antenna (standard antenna). The actual pattern of the antenna could be obtained by comparing the measured data with the data of the standard antenna. At the frequency of 1.25GHz, the actually measured antenna pattern is shown in Fig. 11.

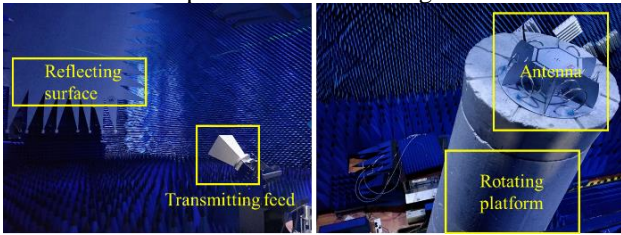

*Fig. 10 Antenna detection platform in microwave anechoic chamber*

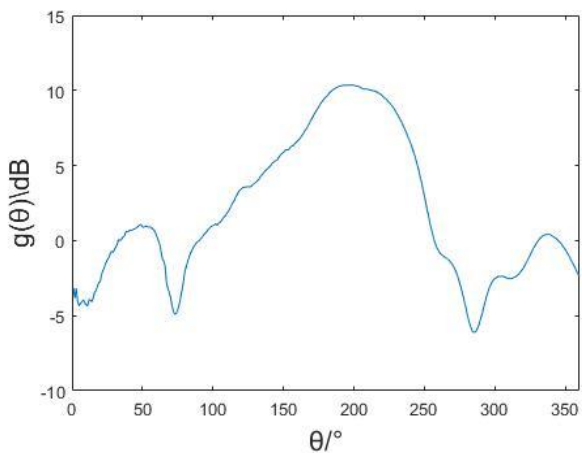

*Fig. 11 the measured pattern*

The result suggests that when the frequency was 1.25GHz, the measured pattern and the simulated pattern had a similar increase and decrease trend. The gain was the largest directly in front of the helical antenna. With the angle shift, the gain gradually decreased, showing a high gain in front of the antenna and a low gain on both sides. The core element of the Dir-MUSIC algorithm based on strength information is that as the direction of the incoming wave changes, the gain of the antenna needs to change drastically, so the designed antenna can meet the requirements of the positioning algorithm.

### 4.2 Positioning Experiment

To verify the reliability of the Dir-MUSIC algorithm based on strength information and the developed antenna, an experimental platform was built as shown in Fig. 12. Based on the conclusion of the previous section, in actual application, if limited by the complexity of sampling equipment and data processing, in an environment with high SNR, using a 4-element array for direction finding could reach the equivalent direction-finding accuracy of a 6-element array. The experimental platform consisted of a digital oscilloscope with storage function, 4 equal-length radio frequency coaxial cables, the developed uniform circular array directional spiral antenna array, and a discharge gun. The location of the experimental site was calibrated by measuring tools such as infrared rangefinders, and PDs are simulated by the discharge gun, performed at different positions of the array. The antenna array received electromagnetic waves and stored them in the digital oscilloscope. Channels 1 to 4 corresponded to the data received by array elements 1 to 4 respectively. Since the volume of the antenna was centimeter-level, its volume was negligible compared to the partial discharge source of 7.5 meters and tens of meters away, so the positions of element 1 to element 4 were (0m, 90°), (0m, 30°), (0m, -30°), (0m, -90°), 20 sets of simulated discharges were performed at the same position.

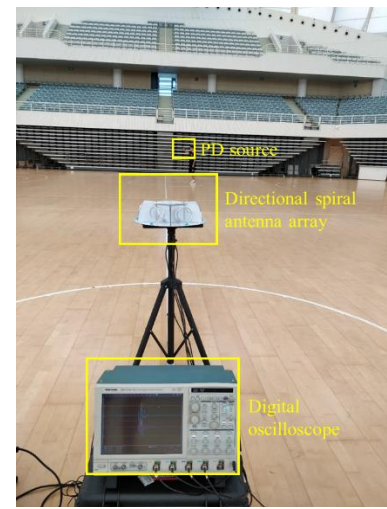

*Fig. 12 Experimental platform*

Discharge was performed at different positions in the experimental area, 20 times at each position. The 4-channel PD pulse waveforms when the discharge position is (7.5m, 10°) are shown in Fig. 12. The width of the discharge pulse waveform was about 2.5ns. Frequency domain analysis on multiple sets of received data was also performed. The energy of the pulse waveform was mainly concentrated between 1.5GHz and 2GHz. There was a maximum value at the frequency of 948MHz in the spectrogram, which was preliminarily presumed to be narrowband interference.

The positioning algorithm flowchart is shown in Fig. 13. Firstly, the array manifold is composed of the measured pattern of a single element. Then based on digital filtering technology, the original data is subjected to band-pass and the pulse waveform area data is intercepted and input into the algorithm module to estimate the direction of the incoming wave. In the algorithm module, the filtered data needs to be mapped to the interval [-1,1] to suppress the dependence of the received signal amplitude on the distance of the discharge source, so that the received data and the array manifold have the same influence on the direction-finding results.

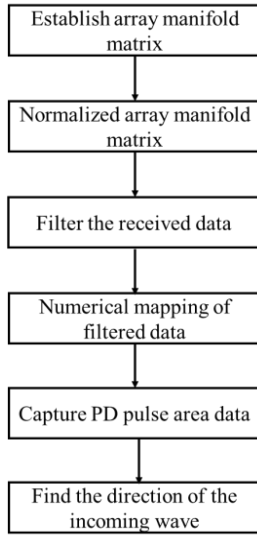

*Fig. 13 Flow chart of Dir-MUSIC algorithm*

Using the Dir-MUSIC algorithm based on strength information to find the direction of 10 sets of experimental data, the results are shown in Table 3. It can be seen from the results that the proposed algorithm had good direction finding performance in multiple incoming wave directions. The maximum mean value of the angle error was 7.55°, the minimum was only 0.7°, and the average multi-direction-finding error was 3.39°. The standard deviation of the angle error was as low as 0.31°, and the maximum was 7.5°. Over 90% of the standard deviation of the error was less than 2.5°, which proved the stability of the algorithm.

**Table 3** Direction finding results of Dir-MUSIC algorithm

| Real PD coordinates | Mean of calculated angle | Mean of angle error | The standard deviation of angle error |
|---|---|---|---|
| (7.5, -96°) | -95.25° | 0.75° | 1.68° |
| (7.5, -66°) | -73.55° | -7.55° | 1.43° |
| (7.5, -19°) | 16.8° | 2.20° | 0.41° |
| (7.5, -10°) | -17.1° | -7.1° | 0.31° |
| (7.5, 10°) | 13.85° | 3.85° | 0.58° |
| (7.5, 17°) | 16.2° | -0.8° | 0.95° |
| (7.5, 28°) | 23.65° | -4.35° | 2.48° |
| (7.5, 93°) | 93.7° | 0.7° | 0.57° |
| (14, -90°) | -93.15° | -3.15° | 2.11° |
| (18.9,17.5°) | 22.05° | 3.45° | 7.5° |

## 5 Conclusion

This paper proposes a novel partial discharge localization Dir-MUSIC algorithm that can be implemented on inspection robots. The method utilizes the strength information of the received signal for positioning, thereby effectively reducing the computational complexity. At the same time, a miniaturized circular array of spiral antennas with 6 elements is developed, and the directivity of the array is calibrated in a microwave anechoic chamber. Finally, the proposed algorithm and the developed array are tested in a real-scene partial discharge environment, and the results proved the effectiveness of the proposed method. The main results and conclusion of this paper are as follows:

(1) Theoretically derived the Dir-MUSIC algorithm based on antenna gain array manifold(AGAM) and signal strength information. Simulated and explored the performance of the algorithm. Through simulation research, it was found that under high SNR, the direction-finding accuracy of the algorithm could be as high as 100%. When the SNR dropped to -10, the direction-finding accuracy could still reach 72.78%. Also，the AGAM error of the algorithm had good robustness.

(2). The optimal array form is determined. Comprehensively considering the direction-finding accuracy, the statistical distribution law of the direction-finding error, the ability of data acquisition processing and array performance, the 6-element uniform circular array layout was the optimal formation, and the direction-finding accuracy of using a 4-element array and a 6-element array was equivalent when the SNR was high.

(3) Carried out the preliminary experimental verification of the sensor system design and Dir-MUSIC algorithm. Based on the positioning requirements, the directional antenna array was researched. The results suggested that the developed directional spiral antenna array pattern was sensitive to angle changes, and its performance met the requirements of the positioning algorithm. The results of the positioning experiment show that the proposed algorithm had good direction-finding performance in multiple incoming wave directions. The average multi-direction-finding error was 3.39°. Also, the dispersion of the direction-finding results was small, which met the positioning requirements for PD localization using inspection robots.

## 6 Acknowledgement

This work is supported by NSFC (51777122) and Natural Science Foundation of Shanghai, China (17ZR1414400).